# High-pressure synthesis, crystal structure determination, and a Ca substitution study of the metallic rhodium oxide $NaRh_2O_4$

*Kazunari Yamaura,[1,*] Qingzhen Huang,[2] Monica Moldovan,[3] David P. Young,[3] Akira Sato,[4] Yuji Baba,[5] Yoshio Matsui,[5] and Eiji Takayama-Muromachi[1]*

Superconducting Materials Center, National Institute for Materials Science, 1-1 Namiki, Tsukuba, Ibaraki 305-0044, Japan;  NIST Center for Neutron Research, National Institute of Standards and Technology, Gaithersburg, Maryland 20899;  Department of Physics and Astronomy, Louisiana State University, Baton Rouge, LA 70803;  Research Promotion Division, National Institute for Materials Science, 1-1 Namiki, Tsukuba, Ibaraki 305-0044, Japan;  Advanced Materials Laboratory, National Institute for Materials Science, 1-1 Namiki, Tsukuba, Ibaraki 305-0044, Japan

TITLE RUNNING HEAD: $NaRh_2O_4$ and $CaRh_2O_4$

* To whom correspondence should be addressed.
E-mail: YAMAURA.Kazunari@nims.go.jp; Fax. +81-29-860-4674

[1] SMC, National Institute for Materials Science
[2] National Institute of Standards and Technology
[3] Louisiana State University
[4] RPD, National Institute for Materials Science
[5] AML, National Institute for Materials Science

ABSTRACT

The sodium rhodate $NaRh_2O_4$ was synthesized for the first time and characterized by neutron and x-ray diffraction studies, measurements of magnetic susceptibility, specific heat, electrical resistivity, and the Seebeck coefficient. $NaRh_2O_4$ crystallizes in the $CaFe_2O_4$-type structure, which is comprised of a characteristic $RhO_6$ octahedra network. The compound is metallic in nature, probably reflecting the 1 to 1 mixed valence character of Rh(III) and Rh(IV) in the network. For further studies of the compound, the Rh valence was varied significantly by means of an aliovalent substitution: the full range solid solution between $NaRh_2O_4$ and $CaRh_2O_4$ was achieved and characterized as well. The metallic state was dramatically altered, and a peculiar magnetism developed in the low Na concentration range.

INTRODUCTION

One of major series of complex oxides in solid state chemistry is the 1:2 compound $AB_2O_4$, which frequently represents a spinel-type group [1]. Regarding the general expression, not only the spinel group, but also other groups are incorporated into this category. For example, another major group in $AB_2O_4$ is the $CaFe_2O_4$-type. The prototype $CaFe_2O_4$ crystallizes in an orthorhombic structure with lattice constants $a = 9.217$ Å, $b = 10.702$ Å, and $c = 3.018$ Å (the space group is *Pnma*) [2], which is built up of eight-fold coordinated Ca atoms and distorted $FeO_6$ octahedra. After the first report of the synthesis of $CaFe_2O_4$ in 1956 [3], a variety of elements were found to replace Ca or Fe, and totally independent A and B elements were found crystallizing in the same structure. As far as we know, A= Ba, Sr, Ca, Mg, Na, La, Eu, and B= La, Pr, Nd, Sm, Eu, Gd, Tb, Dy, Ho, Yb, Lu, Y, Sc, In, Rh, Ti, Fe, V, Cr [2,4], Al [5], Ru [6], Mn [7,8,9], Ga [10], Tl [11] are reported for the $CaFe_2O_4$-type group.

Even though the $CaFe_2O_4$ structure type appears to be rather common, significant attention has not been paid to this group in regards to its magnetic and the electrical properties. There are a few examples: $CaFe_2O_4$ ($Fe^{3+}$: $t_{2g}^3 e_g^2$, $S=5/2$) and $CaMn_2O_4$ ($Mn^{3+}$: $t_{2g}^3 e_g^1$, $S=2$) are anti-ferromagnetic below ~160 K [12] and ~220 K [7,8,9], respectively, and no metallic materials have been discovered thus far, except $NaRu_2O_4$ [6]. $NaTi_2O_4$ is also expected to be a metal from a mixed valence picture of Ti, however, experimental details are unavailable [13,14].

The edge and corner sharing $BO_6$ octahedra in the $CaFe_2O_4$–type structure form a very distinctive network, similar to the one formed in the related perovskite materials. This structural network suggests that interesting physical properties may exist in the $CaFe_2O_4$–type compounds, as are observed in related systems, such as high-$T_c$ superconductivity in cuprates [15], quantum magnetic characters in ruthenates [16], and strongly correlated features in manganates [17]. Motivated by this hypothesis, we have searched a new member of the $CaFe_2O_4$-type group, which has the active $BO_6$ network in a magnetic and electrical sense. Recently, $NaRh_2O_4$ was synthesized by solid-state reaction at a high temperature and pressure. The octahedra network in sodium rhodate contains both $Rh^{3+}$ ($t_{2g}^6 e_g^0$, $S=0$) and $Rh^{4+}$ ($t_{2g}^5 e_g^0$, $S=1/2$) at the 1 to 1 ratio; one half of the $RhO_6$ octahedra are in the spin-1/2 unit.

In this paper, we report synthesis, crystal structure, magnetic and electrical properties of the novel sodium rhodate, and results of studies by aliovalent substitution $Na^{1+}/Ca^{2+}$. A full range solid solution between $NaRh_2O_4$ and $CaRh_2O_4$ [18,19] was synthesized under the same condition.

EXPERIMENTAL

--SAMPLE PREPARATION.

Polycrystalline samples of $Na_{1-x}Ca_xRh_2O_4$ (x=0, 0.25, 0.5, 0.75, 1) were prepared in a platinum cell (6.8 mm in diameter, 0.2 mm in thickness and approximately 5 mm in height) by solid-state reaction in a high pressure apparatus. Fine and pure powders of CaO (99.99 %), $Rh_2O_3$, $NaRhO_2$, and $KClO_4$ (99.5 %) were mixed stoichiometrically, and placed into the platinum cell. Previous to the high-pressure experiment, the $Rh_2O_3$ powder had been prepared from the Rh powder (99.9%) by heating in oxygen at 1000 °C overnight [20]. The powder $NaRhO_2$ had also been synthesized from a mixture of $Na_2O_2$ (98 %) and Rh (99.9 %) powders, which was heated in oxygen at 700 °C for 15 hrs, and then repeating the synthesis at 800 °C for 40 hrs [21]. The platinum cell was heated in the apparatus at 1500 °C for 1 hr at 6 GPa and then quenched at room temperature before releasing the pressure. A technical description of the high-pressure apparatus is detailed elsewhere [22,23]. The sintered samples were dense, black, and retained a pellet shape. Each face of the polycrystalline pellet was polished carefully in order to remove any possible contaminations from chemical reactions with the platinum cell. A typical sample mass was ~0.4 grams.

The samples were examined for quality with powder X-ray (CuKα) diffraction at room temperature on a Rigaku RINT-2000 powder diffractometer, which is equipped with a graphite monochromator on the counter side. In Fig. 1, X-ray diffraction profiles for all the samples are shown. Each peak distribution clearly indicates the $CaFe_2O_4$-type structure is formed in all the samples and the absence of significant impurities except KCl. Details of the qualitative analysis are in a supporting information file (see the end of text). Lattice parameters of the orthorhombic unit cell were determined by a least-squares method, and the evolution of those through the Ca substitution is plotted in Fig. 2. The evolution is small and goes rather monotonically over the Ca/Na range: only approximately 1.7 % change is seen in the unit-cell volume.

Single crystals of $CaRh_2O_4$ were grown in the high-pressure apparatus. Approximately a 0.3-gram mixture of CaO, $CaCl_2$, $Rh_2O_3$, and $KClO_4$ with the ratio 1: 0.109: 0.5: 0.135, respectively, was placed into a double-layered cell consisting of aluminum oxide (inner) and platinum (outer). The sample mixture was isolated from the platinum capsule by the aluminum oxide inner layer. The sample cell was heated at 1500 ºC at 6 GPa for 3 hrs, followed by quenching at room temperature before releasing the pressure. After mechanically removing the layers, a clump of small crystals appeared at the surface of the sample. The crystals were separated by brief treatment in a water sonic bath, and shiny, black crystals up to 0.1 mm in the largest dimension were obtained. The crystal was identified as $CaRh_2O_4$ by an X-ray method (described below). We believe a relatively small degree of temperature gradient in the sample cell between the center and the edge plays a significant role in the crystal growth. Studies to obtain crystals of the other compositions are in progress.

--CHEMICAL ANALYSIS.

A piece of polycrystalline $NaRh_2O_4$ was studied by an energy dispersive X-ray spectroscopy (EDS) at an acceleration voltage of 10 kV in an AKASHI ISI-DS-130 scanning electron microscope. The X-ray spectrum obtained at more than 10 points on the polished surface clearly revealed the absence of platinum contamination. More specifically, platinum was not detected at all above the background level, indicating the platinum concentration was less than 0.1 wt%. We recognized only Na, Rh, O and C (coating material) contributions to the spectra.

In a scanning image at relatively low magnification, a small amount of fragments were found randomly distributed, which exhibited a striking contrast to the major portion. Focused analysis was then conducted on the fragments and revealed the presence of K and Cl. The result is entirely consistent with what was indicated in the X-ray powder diffraction analysis. There are, therefore, no doubts about the formation of KCl residue.

Because we found a small discrepancy in the unit cell parameters between single- and poly-crystal samples of $CaRh_2O_4$; $a = 9.0354(3)$ Å, $b = 3.0340(1)$ Å, $c = 10.7062(3)$ Å, and $V=293.49(2)$ Å$^3$ for single-crystal and $a = 9.035(1)$ Å, $b = 3.081(1)$ Å, $c = 10.78(1)$ Å, and $V=300.0(2)$ Å$^3$ for poly-crystal, a thermogravimetric analysis of the polycrystalline $CaRh_2O_4$ was conducted. The polycrystalline $CaRh_2O_4$ was prepared from CaO and $Rh_2O_3$ without adding $KClO_4$ because the starting mixture was just stoichiometric. Oxygen content of the sample was studied by reduction to calcium monoxide and rhodium by heating in 3%-hydrogen/argon at a heating rate of 2 °C per minute to 800 °C and holding for 8 hours. The measurement was repeated three times. The weight loss data clearly revealed an oxygen-superstoichiometric composition 4.11(3) per formula unit under the synthesis condition, even though the source of the excess oxygen was uncertain. Otherwise, a possible metal deficiency was introduced in the structure as found in the Na-deficient compound $Na_{0.7}(FeAl)_{0.7}Ti_{1.3}O_4$ [24]. For the single crystal, a corresponding non-stoichiometry was not detected in the X-ray refinement study (shown later). The non-stoichiometry in the polycrystalline sample could be responsible for the small discrepancy in the unit cell parameters.

-- ELECTRON DIFFRACTION ANALYSIS.

Selected samples were studied by electron diffraction (ED) on a Hitachi H-1500 electron microscope, in which the electrons were accelerated under a voltage of 820 kV. Careful studies of the compositions $Na_{0.5}Ca_{0.5}Rh_2O_4$ and $Na_{0.25}Ca_{0.75}Rh_2O_4$ by ED revealed that the unit cell was indeed orthorhombic. There were no extra reflections, either sharp or diffuse, which would indicate that there is short range or long range ordering of the Na/Ca site under the conditions of the synthesis. Representative patters are shown in a supporting information file.

--NEUTRON DIFFRACTION ANALYSIS.

Because a single crystal of $NaRh_2O_4$ was thus far unavailable, we decided to conduct a neutron diffraction study on the polycrystalline $NaRh_2O_4$. Approximately 1 gram of powder of the $NaRh_2O_4$ sample was briefly washed in a water sonic bath to remove KCl residue. The powder was then set in the BT-1 high-resolution diffractometer at the NIST Center for Neutron Research, employing a Cu(311) monochromator. Collimators with horizontal divergences of 15', 20', and 7' of arc were used before and after the monochromator, and after the sample, respectively. The calibrated neutron wavelength was $\lambda = 0.15396(1)$ nm, and a drift was negligible during the data collection. Intensity of the reflections was measured at 0.05-degree steps in the 2-theta range between 3 and 168 degrees. The survey was conducted at room temperature. Neutron scattering amplitudes used in data refinements were 0.363, 0.593, and 0.581 ($\times 10^{-12}$ cm) for Na, Rh, and O, respectively.

--SINGLE CRYSTAL X-RAY DIFFRACTION ANALYSIS

A selected $CaRh_2O_4$ crystal was mounted on the end of a fine glass fiber in an area-detector diffractometer (Bruker SMART APEX, MoK$\alpha$ $\lambda=0.71069$ Å). The X-ray study was conducted over night between 25 and 28 °C. Raw data consisting of 1833 frames were collected in $\omega$ scan mode at every 0.3 degrees for 30 seconds ($2\theta_{max}$ was 104.27 degrees). The SMART software was employed for data acquisition and the SAINT+ for data extraction and reduction [25]. An empirical absorption correction was applied with the program SADABS [25]. Structure analysis was attempted on the $F^2$ data by a full-matrix least-squares refinement with the SHELXL-97 program [26].

--PHYSICAL PROPERTIES MEASUREMENTS.

Electrical resistivity of the polycrystalline samples was measured between 2 K and 390 K by a conventional four-point method in a commercial apparatus (Quantum Design, PPMS system). The ac-gage current was 1 mA at 30 Hz. Silver epoxy was used to fix fine platinum wires (~ 30 μm diameter) at four locations along each bar-shaped sample. Thermopower of the samples was measured in the PPMS system between 2 K and 300 K with a comparative technique using a constantan standard. The magnetic susceptibility was measured in a commercial apparatus (Quantum Design, MPMS-XL) at 10 kOe between 2 K and 390 K. The magnetization was studied in the apparatus at 5 K below 70 kOe. Specific-heat measurements were conducted in the PPMS system with a time-relaxation method over the temperature range between 1.8 K and 10 K.

RESULTS AND DISCUSION

Atomic coordination and local structure environment of $NaRh_2O_4$ was investigated by the neutron diffraction study. A Rietveld analysis was applied on the powder diffraction profile with the GSAS program [27]. The structure parameters of $CaFe_2O_4$ were employed as an initial model in the refinement, which aided in obtaining a reliable solution. The best refinement result is shown in Fig. 3. The deference curve (bottom part of the figure) clearly indicates that the raw pattern was precisely reproduced by the model. The refinement details are shown in Tables I and II. In a preliminary study, we temporally unfixed the occupancy factor of Na in order to test for possible non-stoichiometric character. As a result, a very stoichiometric 0.993(24) Na per formula unit was confirmed. We, therefore, decided to fix the Na occupancy - being fully occupied in the final step.

The crystal structure view of $NaRh_2O_4$ was drawn on the basis of the results above (Fig. 4). The structure is nearly identical to that of the prototype compound $CaFe_2O_4$; both have the same space group and similar coordinate environments. The most characteristic feature in the structure should be the double Rh-O chain, which runs along the *b*-axis, as is sketched out at the bottom of Fig. 4. The $RhO_6$ octahedra are connected by edge sharing within each chain, and the chains are tied to neighbors by sharing the corner oxygen. The principal Rh—O—Rh angle in the chain is approximately 98 degrees [average of 97.15(22) and 99.79(16) degrees, see supporting information file]. The intra-chain bond may significantly influence the electrical conductivity (shown later). The inter-chain Rh—O—Rh angles are approximately 130 and 122 degrees. The one dimensional anisotropy of the electronic conducting state might not be expected for this configuration; however it would be interesting to study the degree of the conductivity anisotropy in and out of the chain once a high quality $NaRh_2O_4$ single crystal is available.

The sodium cobalt oxide $NaCo_2O_4$ crystallizes in a layered structure comprised of edge sharing $CoO_6$ octahedra [28]. The layered coordination is believed to play a significant role in the remarkably large thermoelectric power [29]. We were then interested in the thermoelectric properties of $NaRh_2O_4$, because both 1:2 compounds share some common properties, such as the same number of d electrons per B atom and the same structure basis consisting of the edge sharing $BO_6$ octahedra. However, there are structural differences between them. As the most distinguishing structural feature, the layer (NaCo$_2$O$_4$) and chain (NaRh$_2$O$_4$) types, was expected to depend on the relative ionic size of Na and Rh/Co, we considered the Kugimiya and Steinfink (KS) relation for the both compounds [30,31]. The KS relation predicts that the AB$_2$O$_4$ stoichiometry is characterized by two principal parameters: the ratio $r_A/r_B$ (the ionic size factor) and the constant $K_{AB}$ (the bond stretching force factor). The $K_{AB}$ is defined as $K_{AB} = X_A X_B / r_e^2$, where $r_e^2 = (r_A + r_O)^2 + (r_B + r_O)^2 + 1.155(r_A + r_O)(r_B + r_O)$, and $X_A$ ($X_B$) is the electronegativity of the A (B) ion. The $r_A$, $r_B$, and $r_O$ are the radii of A, B, and O ions, respectively. Analyzing the rhodate material first, we obtained $r_A$ = 1.18 Å for Na, $r_B$ = 0.633 Å for Rh, $r_O$ = 1.4 Å, $X_A$ = 1.01 for Na, and $X_B$ = 1.45 for Rh [31,32], and the numerical data yielded the constants $K_{AB}$ = 0.0869 and $r_A/r_B$ = 1.87. Although the result was slightly out of the range of the KS scheme, it is reasonable to include it in the category of the CaFe$_2$O$_4$-type [30]. The result for CaRh$_2$O$_4$ ($K_{AB}$ = 0.0907 and $r_A/r_B$ = 1.68) was within the CaFe$_2$O$_4$-type category and very close to the point for NaRh$_2$O$_4$. The results, therefore, approximates the experimental result quite well. Second, the values for the layered compound NaCo$_2$O$_4$ were also considered, however multiple values of $K_{AB}$ and $r_A/r_B$, which cover all configurations of Co$^{3+}$ and Co$^{4+}$ with low- and high-spin states, indicated that the compound NaCo$_2$O$_4$ was far outside of the range of the KS scheme. It was, therefore, difficult to reach a clear understanding of the role of ionic size in the structural features of NaCo$_2$O$_4$ and NaRh$_2$O$_4$. Further studies will be needed.

To our knowledge, no studies have been reported on the structure of the calcium rhodate CaRh$_2$O$_4$. The structure refinement study was then conducted on a single-crystal of CaRh$_2$O$_4$. The results were compared with those of the prototype CaFe$_2$O$_4$ and the sodium rhodate NaRh$_2$O$_4$. Details of the data collection are in Table III, and the calculated atom positions and thermal parameters are listed in Table IV. A structure distortion in CaRh$_2$O$_4$ was observed. The RhO$_6$ octahedra were found being distorted, but on a relatively small degree; the minimum and maximum distances between Rh and the six ligand oxygen were 1.993(1) Å and 2.054(2) Å, respectively, the difference being ~ 3.1 %. It is obviously smaller than 6.9 % for CaFe$_2$O$_4$ [12] and roughly comparable to 2.1 % for NaRh$_2$O$_4$, and thus suggests keeping the space group as *Pnma*. The structure of $CaTi_2O_4$ is a unique example that crystallizes in a lower symmetry structure than the *Pnma* type [33,34].

Next, oxygen non-stoichiometry was tested and found to be negligible, in striking contrast to the polycrystalline data as already shown (approximately 0.1 mole of excess oxygen was detected). The empirical stoichiometry of the single crystal $CaRh_2O_4$ is identical to that found for $NaRh_2O_4$.

In Fig. 5, temperature and Ca concentration dependence of the electrical resistivity is shown. The resistivity of the title compound $NaRh_2O_4$ is typical of a normal metal. This might reflect the mixed Rh valence character: formally 0.5 unpaired electrons per Rh contribute to the conducting state. As the data indicate, the Ca substitution alters the conducting state dramatically. The state gradually shifts to being poorly conducting with increasing Ca concentration, and the end compound $CaRh_2O_4$ is indeed electrically insulating. The feature is consistent with a simple expectation from the filled $t_{2g}$ band. Considering that the ED observation of the x = 0.5 and 0.75 samples did not indicate any sign of the Na/Ca orderings in either short or long range, the drastic change of over 5 orders of magnitude should reflect the influence of the decreasing carrier density.

The small panel in Fig. 5 shows the same data for $CaRh_2O_4$ in two independent formats, logarithmic $\rho$ vs. $1/T$ (upper, Arrhenius) and $1/T^{0.25}$ (lower, variable range hopping). The data show nearly linear behavior in the latter form, suggesting a hopping conduction mechanism is dominant in $CaRh_2O_4$. This result is in contrast to a preliminary band structure calculation on the basis of the single crystal structure data, which suggests a substantial gap at the Fermi level of approximately 1 eV [35]. Even if the data followed the Arrhenius form, the estimated magnitude of the gap would be ~ 32 meV (as indicated by the dotted line). The inconsistency between the calculation and experiment might result from the non-stoichiometry of the polycrystalline sample, which might supply a small amount of extrinsic carriers and lead to the hopping conduction. The dopant feature was also seen in the magnetic and the thermoelectronic properties of $CaRh_2O_4$ as shown later.

The Seebeck coefficient below 300 K for the solid solution is presented in Fig. 6. The coefficient of the most metallic compound $NaRh_2O_4$ is small and negative, indicating that the transport is dominated by n-type carriers. Unfortunately, the small value of the thermopower precludes the rhodate material for any practical applications, as is found in $NaCo_2O_4$ [28,29]. The Seebeck coefficient increases in magnitude as a function of Ca concentration and reaches a maximum at x = 0.75. Combined with the curvature observed and the zero-crossing, the data suggest that both electron and hole carriers contribute to the transport in this system. The concentration (or mobility) of hole-like carriers apparently increases with increasing Ca content. The two-carrier electronic system and a peak at the $CaRh_2O_4$ composition suggest that single crystal measurements both parallel and perpendicular to the chain direction are needed for a more detailed understanding of the transport properties.

Magnetic susceptibility measured at 10 kOe on cooling is shown in Fig. 7. The fairly metallic compound $NaRh_2O_4$ shows indeed Pauli-type paramagnetism. At room temperature the susceptibility is $\sim 3 \times 10^{-4}$ emu/mol of Rh and decreases with increasing Ca content; this would indicate the density of states at the Fermi level also decreases. The magnetic susceptibility of $CaRh_2O_4$ is fairly small and nearly temperature-independent around room temperature. However, it abruptly rises at low temperature. The feature is most pronounced at x = 0.75 and disappears at the higher Na concentration. A corresponding enhancement is seen in the *M* vs. *H* curve (small panel). The observations are indicative of a possible association between the magnetic enhancement and losing the electrical conductivity. The temperature dependent susceptibility is reminiscent of what is expected for a dilute localized magnetic moment system, as the compounds with low Na content (< ~25 %) are in fact electrically insulating. The small amount of net carriers may be rather localized and help to produce the dilute magnetic moments. Further studies would be required to improve our understanding of the probable correlation between the magnetic and the transport properties of the Na/Ca solid solution.

The specific heat data were quantitatively analyzed in a well-established way. First, the raw data were plotted as $C_p/T$ vs $T^2$ as shown in Fig. 8, and then the following form was applied to fit the linear part by a least-squares method:

$$C_v/T = \gamma + 2.4\pi^4 r N_0 k_B (1/\Theta^3_D) T^2,$$

where $k_B$, $N_0$, and $r$ were the Boltzmann constant, Avogadro's constant, and the number of atoms per formula unit, respectively. The two parameters $\gamma$ (electronic-specific-heat coefficient) and $\Theta_D$ (Debye temperature) are material dependent. The analysis is valid in the low-temperature limit ($T \ll \Theta_D$). The difference between $C_p$ and $C_v$ was assumed insignificant in the temperature range studied. The $\gamma$ and $\Theta_D$ values for $Ca_{1-x}Na_xRh_2O_4$ were then obtained and plotted in the small panel of Fig. 8. The linear part of the data used in the analysis was 20 K$^2$ < $T^2$ < 100 K$^2$ (4.5 K< $T$ < 10 K) for $NaRh_2O_4$ and $Na_{0.75}Ca_{0.25}Rh_2O_4$, 60 K$^2$< $T^2$ < 100 K$^2$ (7.8 K< $T$ < 10 K) for $Na_{0.5}Ca_{0.5}Rh_2O_4$ and $Na_{0.25}Ca_{0.75}Rh_2O_4$, and 3.4 K$^2$< $T^2$ < 100 K$^2$ (1.8 K< $T$ < 10 K) for $CaRh_2O_4$.

The pure Ca sample is electrically insulating, and the $\gamma$ is, therefore, expected to be zero since the Fermi level would lie in the band gap; however, this is not the case. The $CaRh_2O_4$ polycrystalline sample has a $\gamma$ of 2.85(2) mJ/mol of Rh K$^2$. The small but nonzero $\gamma$ suggests, as do the other transport measurements, that a small density of in-gap states exist, probably due to the slight off-stoichiometry of the polycrystalline sample. The small characteristic feature at low temperature at x = 0.75 is indicative of contributions from dilute localized magnetic moments [36,37]. For $NaRh_2O_4$, the $\gamma$ value was 10.13(2) mJ/mol of Rh K$^2$. Using these values of $\gamma$ and $\chi$ (approximately 3×10$^{-4}$ emu/mole of Rh), we find the Wilson ratio for $NaRh_2O_4$ is approximately 2.0 [38], which would suggest the data are somewhat influenced by substantial electron correlations. Further measurements, exploring the correlated behavior in $NaRh_2O_4$ and closely related compounds, may prove very interesting.

In summary, the sodium rhodate $NaRh_2O_4$ was synthesized for the first time under extraordinary conditions. The local structure, magnetic, and electrical properties were studied in detail, and investigation by the aliovalent substitution Na/Ca was conducted. The sodium rhodate $NaRh_2O_4$ was found to be fairly metallic and possibly influenced by substantial electron correlations. The peculiar association of the small amount of electrical carriers and the magnetic moments was suggested for the Na/Ca solid solution. Despite our effort to discover a distinctive feature of the $RhO_6$ network, neither superconductivity nor an ordered magnetic state was observed above 1.8 K below 70 kOe. Further studies on high-quality single crystals would help to clarify and develop our understanding of the physical properties and correlated electron behavior of these compounds. Since intriguing properties due to substantial electron correlations are expected, further synthesis and measurement studies are in progress.

ACKNOWLEDGMENT

We wish to thank M. Akaishi (NIMS) for the high-pressure experiment and H. Aoki (NIMS) for the EDS study. We also express gratitude to Dr. R.V. Shpanchenko for helpful discussion. This research was supported in part by the Superconducting Materials Research Project, administrated by the Ministry of Education, Culture, Sports, Science and Technology of Japan, and by Grants-in-Aid for Scientific Research from the Japan Society for the Promotion of Science (16076209, 16340111).

SUPPORTING INFORMATION PARAGRAPH

Powder x-ray profile of $NaRh_2O_4$ (expanded view), ED patterns, crystallographic information files (CIF), and tables of selected bond distances and angles for $NaRh_2O_4$ and $CaRh_2O_4$. These materials are available free of charge via the Internet at http://pubs.acs.org.

Table I.      Crystallographic data and structure refinement for $NaRh_2O_4$

| | |
|---|---|
| empirical formula | $NaRh_2O_4$ |
| formula weight | 292.798 |
| temperature | 295 K |
| neutron wavelength | 1.5396(1) Å |
| diffractometer | BT-1 at the NIST Center for Neutron Research |
| two theta range used | 3° – 168° in 0.05° steps |
| crystal system | orthorhombic |
| space group | *Pnma* |
| lattice constants | $a$ = 9.0026(4) Å |
| | $b$ = 3.0461(2) Å |
| | $c$ = 10.7268(5) Å |
| volume | 294.16(3) Å$^3$ |
| Z | 4 |
| density (calculated) | 6.611 g/cm$^3$ |
| observations | 2999 |
| R factors | 4.88 % ($R_{wp}$)   3.91 % ($R_p$) |
| Refinement software | GSAS |

Table II. Atomic coordinates and isotropic displacement parameters for NaRh$_2$O$_4$ at 295 K.

| Atom | site | $x$ | $y$ | $z$ | $100U_{iso}$ (Å$^2$) | $n$ |
|------|------|-----|-----|-----|----------------------|-----|
| Na   | 4$c$ | 0.7629(6) | 1/4 | 0.6577(6) | 1.40(12)  | 1 |
| Rh1  | 4$c$ | 0.4137(4) | 1/4 | 0.1029(3) | 0.590(47) | 1 |
| Rh2  | 4$c$ | 0.4431(4) | 1/4 | 0.6162(3) | 0.590(47) | 1 |
| O1   | 4$c$ | 0.1982(4) | 1/4 | 0.1579(3) | 0.759(69) | 1 |
| O2   | 4$c$ | 0.1166(3) | 1/4 | 0.4797(3) | 0.729(77) | 1 |
| O3   | 4$c$ | 0.5319(4) | 1/4 | 0.7862(3) | 0.832(68) | 1 |
| O4   | 4$c$ | 0.4178(4) | 1/4 | 0.4295(3) | 0.598(69) | 1 |

The thermal parameters of Rh atoms were grouped and refined together.

Table III.   Crystallographic data and structure refinement for $CaRh_2O_4$

| | |
|---|---|
| empirical formula | $CaRh_2O_4$ |
| formula weight | 309.90 |
| temperature | 298 K |
| wavelength | MoK$\alpha$ (0.71069 Å) |
| crystal system | orthorhombic |
| space group | *Pnma* (no. 62) |
| unit cell dimensions | $a$ = 9.0354(3) Å |
| | $b$ = 3.0340(1) Å |
| | $c$ = 10.7062(3) Å |
| cell volume | 293.49(2) Å$^3$ |
| Z | 4 |
| density, calculated | 7.014 g/cm$^3$ |
| crystal size (mm) | 0.04×0.02×0.1 |
| *h k l* range | -20$\leq h \leq$18, -5$\leq k \leq$6, -23$\leq l \leq$23 |
| $2\theta_{max}$ | 104.27 |
| linear absorption coeff. | 12.63 mm$^{-1}$ |
| absorption correction | multi-scan (SADABS; Bruker, 1999) |
| $T_{min}/T_{max}$ | 0.5101/0.3815 |
| no. of reflections | 9055 |
| $R_{int}$ | 0.0326 |
| no. independent reflections | 1833 |
| no. observed reflections | 1583 [$F_o$>4$\sigma$($F_o$)] |
| F(000) | 568 |
| R factors | 3.07 % ($R_p$)   8.73 % ($R_{wp}$) |
| weighting scheme | $w = 1/[\sigma^2(F_o^2) + (0.0551P)^2 + 0.02P]$, |
| | $P = (Max(F_o^2) + 2F_c^2)/3$ |
| diff. Fourier residues | [-2.35, 2.66] e/Å$^3$ |
| Refinement software | SHELXL-97 |

Table IV. Atomic coordinates and anisotropic displacement parameters for $CaRh_2O_4$ at 298 K.

---

| Atom | site | $x$ | $y$ | $z$ | $100U_{eq}$ (Å$^2$) | $n$ |
|---|---|---|---|---|---|---|
| Ca  | 4c | 0.23807(8) | 1/4 | 0.34015(7) | 1.228(11) | 1 |
| Rh1 | 4c | 0.08643(3) | 1/4 | 0.60010(2) | 0.569(5)  | 1 |
| Rh2 | 4c | 0.05631(3) | 1/4 | 0.11510(2) | 0.536(5)  | 1 |
| O1  | 4c | 0.3015(3)  | 1/4 | 0.1579(3)  | 0.702(28) | 1 |
| O2  | 4c | 0.3825(2)  | 1/4 | -0.0236(2) | 0.526(26) | 1 |
| O3  | 4c | 0.4688(3)  | 1/4 | 0.2114(2)  | 0.598(27) | 1 |
| O4  | 4c | 0.0850(2)  | 1/4 | -0.0726(2) | 0.528(27) | 1 |

| Atom | $100U_{11}$ | $100U_{22}$ | $100U_{33}$ | $100U_{23}$ | $100U_{13}$ | $100U_{12}$ |
|---|---|---|---|---|---|---|
| Ca1 | 0.72(2)  | 2.16(3)  | 0.808(19) | 0 | 0.004(16) | 0 |
| Rh1 | 0.514(8) | 0.677(8) | 0.515(7)  | 0 | -0.016(5) | 0 |
| Rh2 | 0.512(8) | 0.553(8) | 0.543(7)  | 0 | -0.022(5) | 0 |
| O1  | 0.35(6)  | 0.76(8)  | 1.00(7)   | 0 | 0.03(6)   | 0 |
| O2  | 0.33(6)  | 0.77(7)  | 0.48(6)   | 0 | 0.02(5)   | 0 |
| O3  | 0.60(7)  | 0.75(7)  | 0.45(6)   | 0 | -0.15(5)  | 0 |
| O4  | 0.37(6)  | 0.56(7)  | 0.65(6)   | 0 | -0.04(5)  | 0 |

Figures and figure captions

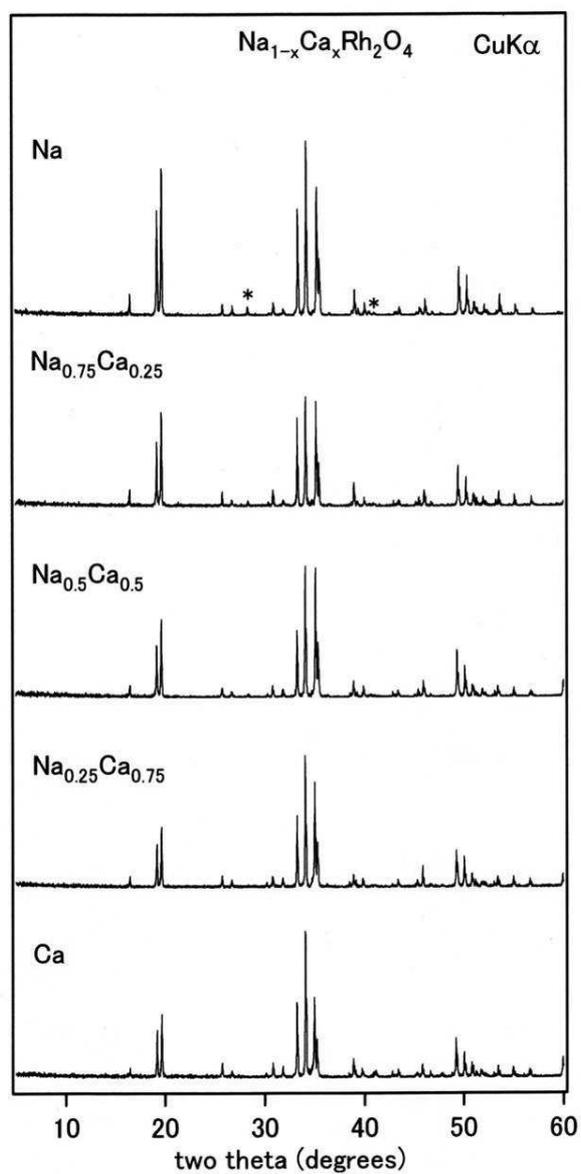

Fig1. Powder x-ray diffraction profile of the $Na_{1-x}Ca_xRh_2O_4$ samples, measured at room temperature. Star marks indicate peaks for KCl. For clarity *hkl* indexes are not shown for here, but an expanded view with the indexes is presented in a supporting information file.

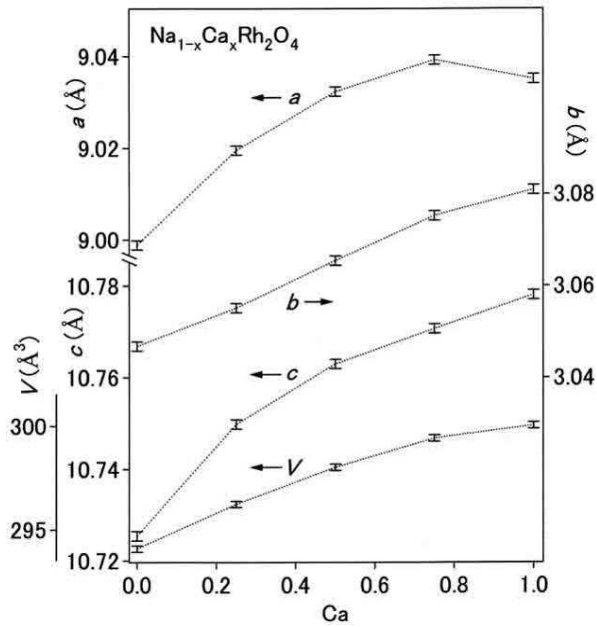

Fig2. Orthorhombic unit-cell parameters and the unit-cell volume of $Na_{1-x}Ca_xRh_2O_4$, measured at room temperature by an X-ray diffraction method.

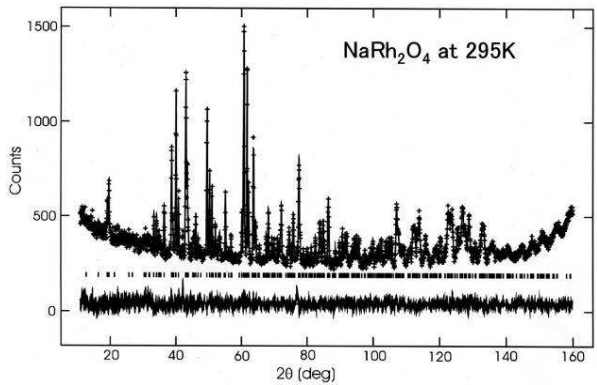

Fig3. Neutron diffraction profile of the $NaRh_2O_4$ sample (~1 gram), measured at 295 K. Vertical bars indicate allowed Bragg reflections on the basis of the *Pnma* structure. The difference between the best computed profile (solid lines) and the raw data (crosses) is shown below the bars column.

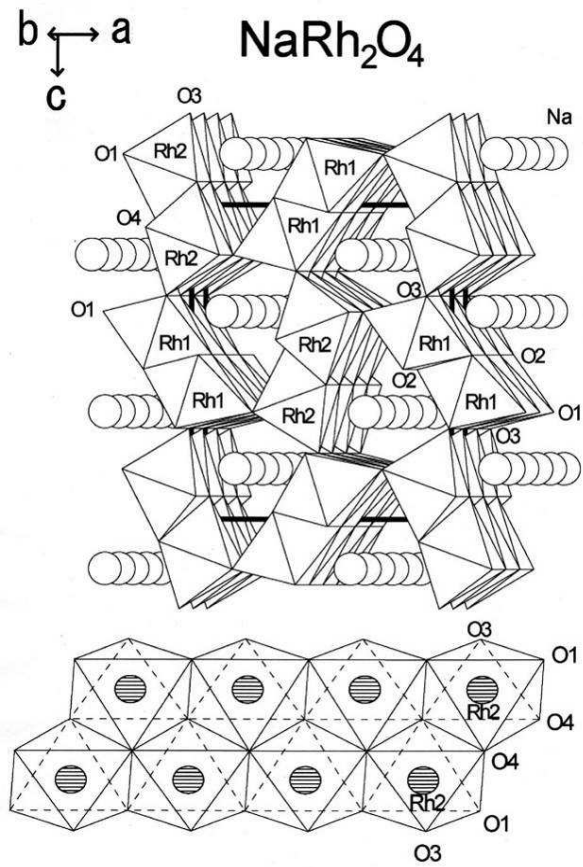

Fig4. Structure view of NaRh$_2$O$_4$. Bold lines signify the orthorhombic unit cell. Bottom figure indicates a part of the double chain along the *b*-axis.

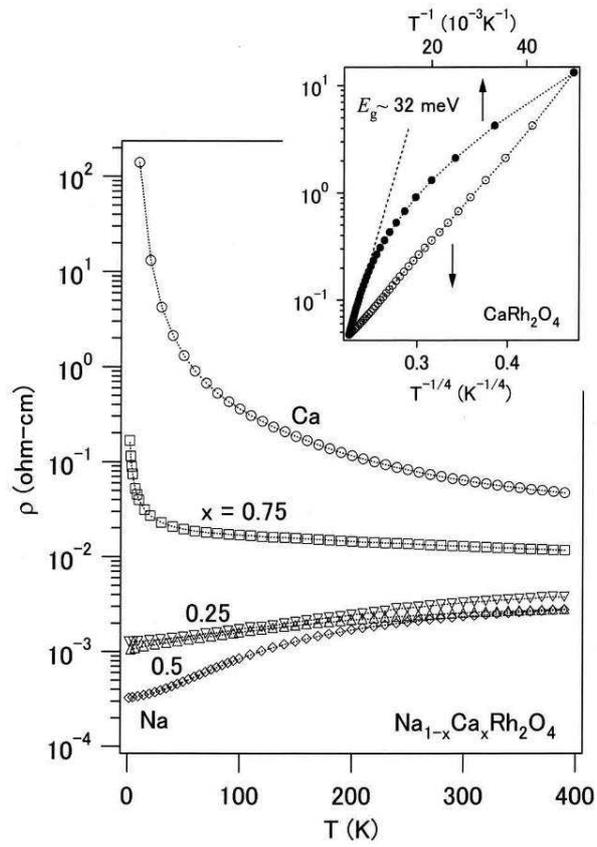

Fig5. Temperature and composition dependence of the electrical resistivity of the polycrystalline $Na_{1-x}Ca_xRh_2O_4$. (Top panel) Comparison between the two plots of the data for $CaRh_2O_4$.

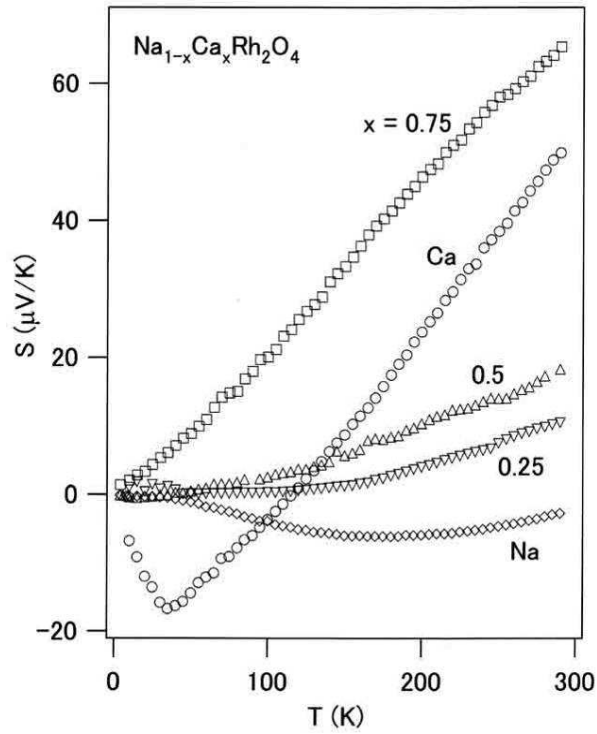

Fig6. Thermoelectric power of the polycrystalline $Na_{1-x}Ca_xRh_2O_4$.

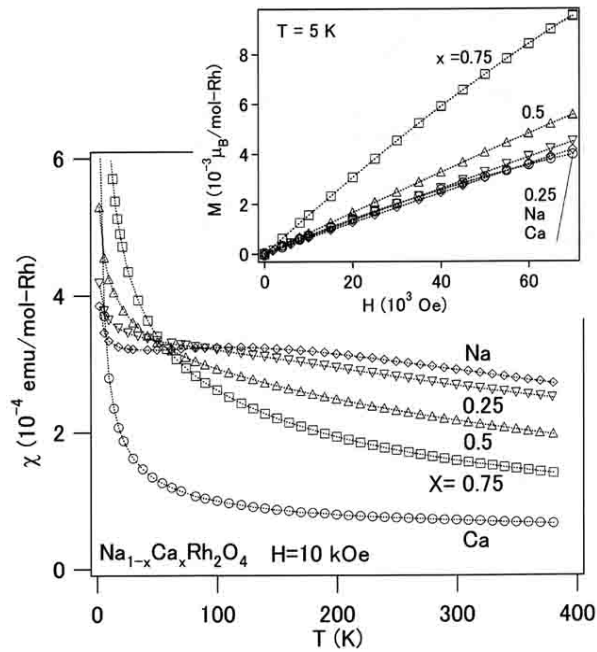

Fig7. Temperature dependence of the magnetic susceptibility of the polycrystalline $Na_{1-x}Ca_xRh_2O_4$, measured at 10 kOe on cooling, and applied field dependence of the magnetization at 5 K (top panel).

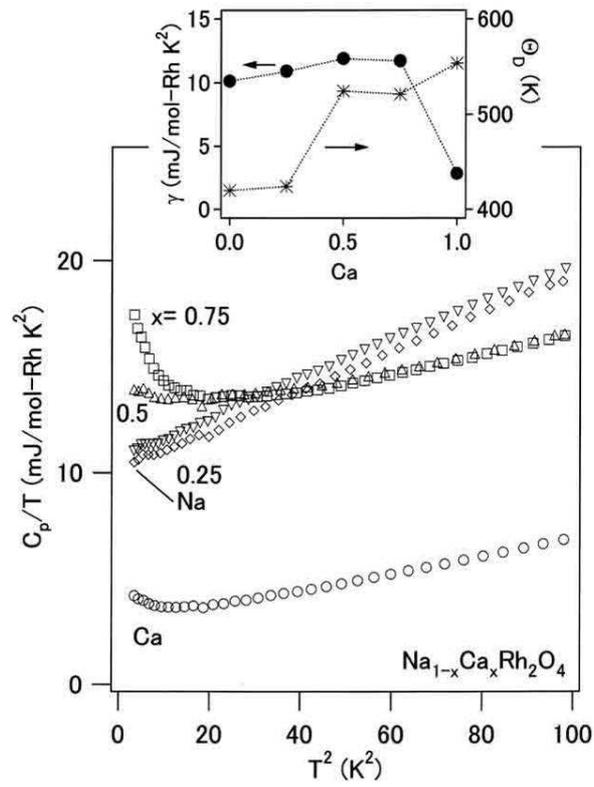

Fig8. Specific heat of the polycrystalline $Na_{1-x}Ca_xRh_2O_4$. Small panel shows evolution of $\gamma$ (electronic-specific-heat coefficient) and $\Theta_D$ (Debye temperature), estimated from the data. Error bars of the $\gamma$ and $\Theta_D$ points are smaller than the marker size.